\documentclass[%
 reprint,
 amsmath,amssymb,
 aps,
]{revtex4-2}
\usepackage{tikz}
\usepackage{subfigure}
\usepackage{dcolumn}% Align table columns on decimal point
\usepackage{braket} % For Dirac notation like \ket
\usepackage{float}      % for the [H] placement specifier
\usepackage{bm}% bold math
\usepackage{mathrsfs}
\usepackage{placeins}
\begin{document}
\newcommand{\hrho}{\ensuremath{\mathscr{H}_{\rho}}}
\newcommand{\hrhoinv}{\ensuremath{\mathscr{H}_{\rho^{-1}}}}
\newcommand{\h}{\ensuremath{\mathscr{H}}}
\newcommand{\pt}{\ensuremath{\(\mathcal{PT}\)}}
\newcommand{\diff}{\ensuremath{\mathrm{d}}}

\preprint{APS/123-QED}

\title{Emulation of Self-Consistent Non-Hermitian Quantum Formalisms}% Force line breaks with \\
%\thanks{Shorter title}%

\author{Mario Gonzalez}
% \email{mormario@student.ethz.ch}
\affiliation{Institute for Theoretical Physics, ETH Z\"{u}rich, 8093 Zurich, Switzerland}%Lines break automatically or can be forced with \\

\author{Karin Sim}
\email{simkarin@phys.ethz.ch}
 % \homepage{http://www.Second.institution.edu/~Charlie.Author.}
\affiliation{Institute for Theoretical Physics, ETH Z\"{u}rich, 8093 Zurich, Switzerland}

\author{R. Chitra}%
 \email{chitrar@ethz.ch.}
\affiliation{Institute for Theoretical Physics, ETH Z\"{u}rich, 8093 Zurich, Switzerland}

\date{\today}% It is always \today, today,

\begin{abstract}

%Non-Hermitian quantum mechanics is an emerging frontier of physics  with potential to redefine standard paradigms of  physics.
%, harboring a plethora of unconventional phenomena with no Hermitian counterpart, such as faster-than-Hermitian information transfer, exotic entanglement effects and non-Hermitian skin effect. 

%Non-Hermitian quantum mechanics is an emerging frontier of physics  with potential to redefine standard paradigms of  physics.
%, harboring a plethora of unconventional phenomena with no Hermitian counterpart, such as faster-than-Hermitian information transfer, exotic entanglement effects and non-Hermitian skin effect. 
Standard quantum mechanics  predicts  the non-conservation of state norms and probability when the fundamental requirement of the Hermiticity of the Hamiltonian is relaxed. Biorthogonal quantum mechanics, or the more general metric formalism, provides  a rigorous formulation of non-Hermitian quantum mechanics  wherein  norms and probabilities are conserved.  The key feature is  that the Hilbert space is endowed with a non-trivial dynamical metric.  
%Non-Hermitian dynamics as approximations to open quantum systems   conform  to the standard quantum physics paradigm.  
Beyond theoretical considerations, the physical implementation of the metric formalism remains unaddressed.
 In this work, we propose novel operator dilation schemes, which show that  the self-consistent non-Hermitian quantum mechanics can be accessed in physical platforms via an embedding in closed Hermitian systems.
Using digital quantum simulators, we present  a  proof of principle and the first experimental  evidence for   the dynamical metric  engendered by  non-Hermiticity in  a qubit.  Our work ushers  in a new paradigm in the quantum simulation of non-Hermitian systems.

\end{abstract}

\maketitle

%\tableofcontents

% \section{\label{sec:introduction}Introduction}

\textit{Introduction.}--- Hermiticity of the Hamiltonian lies at the core of standard quantum mechanics. This allows for the interpretation of the Hamiltonian as an energy observable and ensures unitary time evolution.
Under this assumption, both the inner product structure and the norms of states in the Hilbert space are preserved in time, allowing for a probabilistic interpretation of quantum mechanics\,\cite{dirac1958principles}. 
In the past decade, there has been extensive focus on  non-Hermitian Hamiltonians \cite{bender1998real,bender2007making,mostafazadeh2002pseudo,mostafazadeh2004physical,mostafazadeh2006physical,mostafazadeh2007quantum,mostafazadeh2010pseudo,mostafazadeh2020time}.
A non-Hermitian Hamiltonian which commutes with the   parity-time-reversal operator  $(\mathcal{PT})$ can possess a real spectrum\,\cite{bender1998real,streater2000pct}, while exceptional points  and complex eigenvalues emerge in the spectra 
when this $\mathcal{PT}$-symmetry is spontaneously broken\,\cite{bender2007making,neumark1940spectral,naimark_science}.  These features lead to a  plethora of rich physics with no Hermitian counterparts \cite{ashida2020non,bender2007faster,zhan2020experimental,turkeshi2023entanglement,couvreur2017entanglement}. 
 % can however, be spontaneously broken when the eigenstates of the Hamiltonian are no longer eigenstates of the $(\mathcal{PT})$ operator. 

The true richness of non-Hermitian quantum mechanics unveils in the time domain.
As state norms are not preserved during non-unitary time evolution,   
to obtain well-defined expectation values, the oft-used approach is to simply normalize the time-evolved states with respect to their time-dependent norms. For the sake of clarity, we henceforth refer to this approach as the {\it norm method}. This approach correctly describes the non-Hermitian dynamics of quantum trajectories without quantum jumps in open quantum systems. It leads to 
 unconventional phenomena, such as faster-than-Hermitian evolution\,\cite{bender2007faster,mostafazadeh2007quantum}, violation of Lieb-Robinson bounds for information propagation\,\cite{zhang2024observation} and quantum cloning\,\cite{zhan2020experimental} to name a few.
\,\cite{shibata2019dissipative,martinez2024quantum,lu2024realizing,budich2020non,mcdonald2020exponentially,wiersig2020review,schomerus2020nonreciprocal}. %, where the quantum metric does not play a role in the dynamics. 

An alternative self-consistent framework, the metric formalism\,\cite{mostafazadeh2004physical,mostafazadeh2010pseudo,hornedal2024geometrical,mostafazadeh}, allows for a consistent definition of probabilities and expectation values which we briefly summarize below. This is a generalization of the better known biorthogonal quantum mechanics\,\cite{brody2013biorthogonal}.
For a system described by a general non-Hermitian Hamiltonian $H(t)$, the metric framework postulates the emergence of a modified Hilbert space \((\hrho, H(t))\) with a consistently determined dynamical inner product weighted by the metric \(\langle\cdot,\cdot\rangle_{metric}:=\langle\cdot|\rho(t)|\cdot\rangle\) \,\cite{mostafazadeh2004physical,mostafazadeh2010pseudo,hornedal2024geometrical,mostafazadeh} .
This restores unitary evolution within the modified Hilbert space, resulting in a consistent probabilistic interpretation.
The metric $\rho(t)$ is a time-dependent operator which evolves according to\,\cite{mostafazadeh}
\begin{equation}
    i\dot{\rho}(t)=H^\dagger(t)\rho(t)-\rho(t)H(t),
    \label{eq:tdqh}
\end{equation}
which incorporates the dynamics of both $H(t)$ and $H^\dagger(t)$.
We note that $\rho(t)$ is neither a metric in the strict sense of a map in a metric space nor does it correspond to the quantum geometric tensor discussed in Refs.\,\cite{tensor1,tensor2}. The metric, however, cannot be directly accessed in a closed non-Hermitian system\,\cite{brody2016consistency}, as the metric is part of the inner product and an observer could as well describe his closed system with a Hermitian counterpart.
Nevertheless, the metric dynamics drastically alters the physics, as evinced by the restoration of information bounds in the quantum Brachistochrone problem\,\cite{mostafazadeh2007quantum} and the violation of quantum adiabaticity via defect freezing of $\mathcal{PT}$-broken modes\,\cite{defectfreezing}. These studies show that this self-consistent formulation of quantum mechanics, encapsulated by the metric, harbors a phenomenological richness yet to be explored.

An intriguing question is whether the metric formalism is purely of theoretical interest or  can it be realized in an experiment, especially when nature is described by effectively closed Hermitian systems or open systems. 
In this work, we provide a pathway to realizing the metric framework of non-Hermitian quantum mechanics and obtain the first direct experimental measurement of the dynamical metric alongside other observables. 
Using operator dilation theory\,\cite{shalit2021dilation},  we present two new dilation schemes that embed the  non-Hermitian system of interest  in  a larger Hermitian system. 
Our schemes encompass both norm  and the metric formalisms,   each of  which can be directly accessed via projective measurements.
We  implement our dilation protocols for a two level non-Hermitian system on IBM Quantum’s superconducting backend \textbf{ibm\_Kyiv} accessed via Qiskit. 
Using tomographic reconstructions, our results perfectly capture the metric dynamics, which is a key ingredient in the self-consistent framework of non-Hermitian quantum mechanics.
%\textcolor{blue}{Our results (Change wording, results appears 3 times)} \textcolor{olive}{highlight} that the metric is a time periodic operator in the $\mathcal{PT}$-symmetric phase \textcolor{olive}{while in the $\mathcal{PT}$-broken region the behavior becomes hyperbolic.} 
% \textcolor{blue}{Put this somewhere: Our results perfectly capture the metric dynamics, which is a key ingredient in the self-consistent framework of non-Hermitian quantum mechanics.} 

\emph{The metric formalism.}---Assume a general non-Hermitian Hamiltonian $H(t)$ and its Hermitian metric operator $\rho(t)$. 
We first note that via the factorization into its principal square root $\rho(t)=\eta^2(t)$\,\cite{mostafazadeh2020time,defectfreezing}, $\rho(t)$ dictates a mapping to an equivalent Hermitian system in terms of a conventional Hilbert space \((\h, h(t))\).
This Hilbert space features a standard inner product \(\langle\cdot,\cdot\rangle:=\langle\cdot|\cdot\rangle\), and the dynamics is governed by a Hermitian Hamiltonian $h(t)$ 
% The metric is positive definite by construction, therefore one can always diagonalize it and find the unique principal root \textcolor{blue}{[$S^{-1}$ or $S^\dagger$?]}\textcolor{red}{[True, I think you are right because $\rho(t)=\rho^\dagger(t)$]}$\rho(t)=S\Lambda S^\dagger =S\Lambda^{1/2} S^\dagger S\Lambda^{1/2} S^\dagger$, where $\Lambda$ is a diagonalized operator in a basis defined by the unitary $S$. 
% This implies that $\eta(t)=S\Lambda^{1/2} S^{-1}$ is unique \textcolor{blue}{[check]}.
defined by
\begin{equation}
    h(t)=\eta(t)H(t)\eta^{-1}(t)+i\dot{\eta}(t)\eta^{-1}(t).\label{eq:H_h_mapp}
\end{equation}
The state evolution in both Hilbert spaces is governed by the Schr\"odinger equation:
$i|{\dot \psi}(t)\rangle=H(t)|\psi(t)\rangle$ and 
 $i|{\dot \Psi}(t)\rangle=h(t)|\Psi(t)\rangle$,
where the time-evolved states are related by $|\Psi(t)\rangle=\eta(t)|\psi(t)\rangle$. The state evolved with the Hermitian conjugate, $H^\dagger(t)$, is related to $\ket{\psi(t)}$ by the metric, i.e. $i\frac{d}{d t}(\rho(t)|\psi(t)\rangle)=H^\dagger(t)(\rho(t)|\psi(t)\rangle)$. 
In this formalism, the expectation value of a general observable $O$ is defined as $\langle O\rangle_{m}(t)=\bra{\psi(t)}\eta(t)O\eta(t)\ket{\psi(t)}$.
% These two descriptions, which we refer to as the \textbf{metric} method, are equivalent to each other, but the physical interpretations differ. 
%\textcolor{blue}{which incorporates both $H(t)$ and its Hermitian conjugate $H^\dagger(t)$ in its evolution(?)}.

In this work, we set the initial condition to be $\rho(t_0)=\mathbb{I}$, since we assume that the starting state can be reached via Hermitian dynamics. 
The metric can be written as $ \rho(t)=U_{H^\dagger}(t,t_0)U_H(t_0,t)$, where $U_H(t,t_0)$ and $U_{H^\dagger}(t,t_0)$ are the time evolution operators for $H(t)$ and $H^\dagger(t)$, respectively i.e.,
\begin{equation}
  U_A(t,t_0) = \mathscr{T}(e^{-i\int_{t_0}^t dt A(t)}),
\end{equation}
where $\mathscr{T}$ is the time-ordering operator. 
Clearly, $\rho(t)=\mathbb{I}$ iff $H^\dagger(t)=H(t)$, implying that the definition of the dynamical norm reduces to the standard norm, \(\langle\cdot|\rho(t)|\cdot\rangle=\langle\cdot|\cdot\rangle\) for Hermitian Hamiltonians.  

% {\color{blue}  U not explicitly defined ......mention usual time evolution operator}
%\textcolor{blue}{[A couple of sentences to introduce PT symmetry and its significance (they possess real spectra, can therefore be mapped to Hermitian Hamiltonians via similarity transformations)]} 
% \textcolor{red}{[Actually, for the content of the paragraph below, you can just refer to the second paragraph of https://journals.aps.org/prresearch/abstract/10.1103/PhysRevResearch.7.013325]}\\

% \textcolor{red}{[Add a few sentences about the spontanteous breaking of the PT symmetry and the exceptional point (where the eigenvectors coalesce). When PT-symmetry is spontaneously broken, the eigenstates of the Hamiltonian are no longer eigenstates of the PT operator. The energy goes from real to imaginary beyond the EP. (This is important for later as we distinguish the behaviour between pt-symmetric and pt-broken regimes)]}
% Consider adding the following references for similarity transformation (check for duplicates): Ashok Das 2011 J. Phys.: Conf. Ser. 287 012002 // Dorje C Brody 2014 J. Phys. A: Math. Theor. 47 035305 // PSEUDO-Hermitian REPRESENTATION OF QUANTUM MECHANICS, International Journal of Geometric Methods in Modern Physics 2010 07:07, 1191-1306 // J. Math. Phys. 43, 205–214 (2002)
For a constant, $\mathcal{PT}$-symmetric Hamiltonian with real eigenvalues, a constant solution to Eq.\,\eqref{eq:tdqh} exists\,\cite{sim2025observables,mostafazadeh2010pseudo,brody2013biorthogonal,das2011pseudo}, where $\rho_CH=H^\dagger\rho_C$ and $\dot{\rho}_C=0$. 
% {\color{blue} remove:In other words, this constant solution is the similarity transformation which maps $H$ to its Hermitian conjugate $H^\dagger$\,.}
However,  other initial conditions,   including $\rho(t_0)=\mathbb{I}$, imposes a dynamical metric.
%Note, however, that one can also find so called similarity transformations $\rho_c$ which are constant solutions to the tdqH equation: . The choice of the metric is therefore not unique. 

%\textcolor{blue}{[Up to this point, you're only doing a literature survey, recapping what is in the literature and where there's a gap. Let's keep it this way and not touch on too much about what we've done.]}
%\textcolor{blue}{[From this point on,  we switch gears and talk about what we have done - so I'd add a paragraph here and put a couple of sentences to introduce our goal (propose a novel scheme to simulate the physics encapsulated by metric). In PRL, you're catering to a broad community. It's not immediately clear what a `dilation' is to people from other communities... So in the paragraph, I'd also explain why the digital simulator is a suitable platform, and why we need to embed  a non-Hermitian system in a larger Hermitian system (since we only have unitary gates to work with),  what has been done before (Naimark), and the problems it faces (not always valid in the PT-broken regime).]} 

To be able to observe and implement the metric, we need to move beyond the notion of a closed system. This can be done using operator dilation theorem\,\cite{shalit2021dilation}, which permits the implementation of non-unitary evolutions with the help of an ancilla system and projective measurements.
%\textcolor{blue}{[I'd move this paragraph to the paragraph before Eq.(7) (see my comment there).]} 
An example of a dilation scheme is the Naimark dilation\,\cite{ naimark_science,dogra2021quantum,beneduci2020notes,lu2024realizing}, an established method in quantum information, where
the total system-ancilla state takes the form 
\begin{equation}
    \ket{\Psi_N(t)}=\ket{\psi(t)}_s\otimes\ket{0}_a+\tilde{\eta}(t)\ket{\psi(t)}_s\otimes\ket{1}_a
    \label{eqn:naimarkstate}
\end{equation}
with $\tilde{\eta}(t)=\sqrt{\rho(t)-\mathbb{I}}$.
% {\color{blue}  ancilla states missing in the above eqn}
Projecting onto the $\ket{0}_a$ state of the ancilla, one realizes the non-unitary evolution of $\ket{\psi(t)}_s$ corresponding to the usual norm method for non-Hermiticity.
This dilation has been implemented both in solid state and quantum circuit settings\,\cite{lu2024realizing, naimark_science,dogra2021quantum}. 
However, this implementation does not capture the metric dynamics as $\tilde{\eta}(t)\ket{\psi(t)}$ cannot be straightforwardly mapped to the dynamics of the Hamiltonian $h(t)$.  
Furthermore,  $\tilde{\eta}(t)$, and consequently Eq.\,\eqref{eqn:naimarkstate}, are generally ill-defined beyond a certain time, as $\rho(t)-\mathbb{I}$ is not guaranteed to be positive semi-definite in the $\mathcal{PT}$-broken regime. The latter is attributable to the exponential decay of at least one of the eigenvalues of $\rho(t)$ or $\rho^{-1}(t)$ with time \,\cite{zhang2019time}.

\begin{figure}[!t]
\centering
    \begin{minipage}{0.9\linewidth}
    \centering
    \begin{tikzpicture}
        \node[anchor=north west] at (-0.2,1.5) {\small (a)};
        % \draw[purple, thick, fill=purple!20] (1.5, -0.5) circle (1.5);
        
        \draw[thick, fill=red!20, rounded corners=3pt] (0, -1) rectangle (2, 1); %node[midway] {$\rho(t)\ket{\psi(t)}$}; 

        \draw[thick, fill=blue!20, rounded corners=3pt] (2.25, -1) rectangle (4.25, 1); % node[midway] {$\ket{\psi(t)}$}; 

        \draw[thick, fill=green!20, rounded corners=3pt] (4.5, -1) rectangle (6.5, 1); %node[midway] {$\eta(t)\ket{\psi(t)}$}; 
        
        % Arrows between boxes
        % \draw[blue,->, thick, bend right=40] (1, -1) to (2.25, -1); 
        
        % \draw[red,->, thick, bend right=40] (2.25, 0) to (1, 0); 
        
        % \draw[red,->, thick, bend right=30] (4, 0) to (2.25, 0); 

        % \draw[blue,->, thick, bend right=30] (2.25, -1) to (4, -1); 

        % Nodes above and bellow arrows

        % \node[blue] at (1.625, -1.5) {\textbf{ $\rho^{-1}(t)$}};

        % \node[blue] at (3.125, -1.5) {\textbf{ $\eta(t)$}};

        % \node[red] at (1.625, 0.5) {\textbf{ $\rho(t)$}};

        % \node[red] at (3.125, 0.5) {\textbf{ $\eta^{-1}(t)$}};

        % Nodes for the different colors
        \node[red!50!black] at (1,0.75) {\textbf{$(\hrhoinv,H^\dagger)$}};

        \node[blue!50!black] at (3.25,0.75) {\textbf{$(\hrho,H)$}};

        \node[green!50!black] at (5.5,0.75) {\textbf{$(\h,h)$}};

        % Nodes and arrows for the states
        \node[red!50!black] at (1,0.25) {\textbf{$\ket{\psi(t_0)}$}};

        \draw[red,->, thick] (1, 0) to (1, -0.5); 

        \node[red!50!black] at (1,-0.75) {\textbf{$\rho(t)\ket{\psi(t)}$}};

        \node[blue!50!black] at (3.25,0.25) {\textbf{$\ket{\psi(t_0)}$}};

        \draw[blue,->, thick] (3.25, 0) to (3.25, -0.5); 

        \node[blue!50!black] at (3.25,-0.75) {\textbf{$\ket{\psi(t)}$}};

        \node[green!50!black] at (5.5,0.25) {\textbf{$\ket{\psi(t_0)}$}};

        \draw[green!25!black,->, thick] (5.5, 0) to (5.5, -0.5); 

        \node[green!50!black] at (5.5,-0.75) {\textbf{$\eta(t)\ket{\psi(t)}$}};
    \end{tikzpicture}
    \end{minipage}
    \\
    \begin{minipage}{0.9\linewidth}
    \centering
    \begin{tikzpicture}
    \node[anchor=north west] at (-1.4,1.2) {\small (b)};
    % Draw the quantum wires
    \draw[thick] (-0.5, 0) -- (4.25, 0) node[anchor=west] {\(\)};
    \draw[thick] (-0.5, -0.5) -- (4.25, -0.5) node[anchor=west] {\(\)};
    \draw[thick] (-0.5, -1) -- (4.25, -1) node[anchor=west] {\(\)};

    % Initial state text fields
    % \node[anchor=east] at (-0.5, 0) {$\ket{\psi(t_0)}$};
    \node[anchor=east] at (-0.5, 0) {$\ket{\psi(t_0)}$};
    \node[anchor=east] at (-0.5, -0.5) {$\ket{+}_a$};
    \node[anchor=east] at (-0.5, -1) {$\ket{+}_{a'}$};

    % Two-qubit gate
    \draw[thick, fill=yellow!20, rounded corners=3pt] (2.65, -1.3) rectangle (3.75, 0.3) node[midway] {$\zeta_G^{-1}(t)$};

    % Two-qubit gate
    \draw[thick, fill=yellow!20, rounded corners=3pt] (0, -1.3) rectangle (1.1, 0.3) node[midway] {$\zeta_G(t_0)$};

    % Single-qubit gate
    \draw[thick, fill=yellow!20, rounded corners=3pt] (1.25, -0.3) rectangle (2.5, 0.3) node[midway] {$U_h(t,t_0)$};

    % Rectangle around gates with text
    \draw[red, thick, dashed, rounded corners=5pt] (-0.25, -1.5) rectangle (4, 0.5);
    \node[anchor=south] at (1.75, 0.5) {$U_{tot}(t,t_0)$};

    % Final entangled state text field
    % \node[blue!50!black,anchor=west] at (4.25, 0) {$\ket{\psi(t)}\otimes\ket{00}+$};

    \node[black!50!black,anchor=west] at (4.25, -0.5) {$\Bigg\}\ket{\Psi(t)}$};

\end{tikzpicture}
\end{minipage}
\\
\begin{minipage}{0.9\linewidth}
\centering
    \begin{tikzpicture}
    \node[anchor=north west] at (-2.4,1) {\small (c)};
    \draw[thick, fill=blue!20, rounded corners=3pt] (-0.15, 0.6) rectangle (0.7, 0.1) ;

    \draw[thick, fill=green!20, rounded corners=3pt] (-0.15, -0.1) rectangle (1.35, -0.6) ;

    \draw[thick, fill=red!20, rounded corners=3pt] (2, 0.6) rectangle (3.45, 0.1) ;
    \node[black!50!black,anchor=west] at (-0.25, 0.35) {$\ket{\psi(t)}\otimes\ket{00}+\rho(t)\ket{\psi(t)}\otimes\ket{10}+$};

    \node[black!50!black,anchor=west] at (-0.2, -0.35) {$\eta(t)\ket{\psi(t)}\otimes(\ket{01}+\ket{11})$};
    
    \node[black!50!black,anchor=west] at (-2.95, 0) {$\ket{\Psi(t)}=\frac{1}{\sqrt{C(t)}}\Bigg($};

    \node[black!50!black,anchor=west] at (4.65, 0) {$\Bigg)$};
    
\end{tikzpicture}
\end{minipage}
\caption{  (a) Schematic of the Hilbert spaces with their corresponding states evolving according to $H(t)$, $H^\dagger(t)$ and $h(t)$, respectively. The metric $\rho(t)$ and its principal root $\eta(t)$ serve as the mappings between the states. (b) The circuit layout of the Generalized Biorthogonal Naimark dilation (GBoNd), which consists of a system qubit and two ancilla qubits. This protocol can be understood as a unitary evolution $U_h(t, t_0)$ sandwiched by non-trivial dilations $\zeta_G(t)$ of $\rho_G^{1/2}(t)=\eta_G(t)$ as defined in Eq. (\ref{eqn:zetag}). (c) The final state $\ket{\Psi(t)}$ obtained in the GBoNd protocol, where $C(t)$ is a normalization factor.  
The colors of the boxes indicate the components of the total states evolving under different Hamiltonians, as shown in (a).
In the postselection after measuring the ancilla states, the effectively normalized states are measured: $\ket{\psi(t)}/|\psi(t)|$, $\rho(t)\ket{\psi(t)}/|\rho(t)\psi(t)|$ and $\eta(t)\ket{\psi(t)}$, with the third being normalized by construction.}
% \textcolor{red}{[1. use fancy $\h$ and $\hrho$ instead of mathcal- check paragraph above Eq.(1) 2. I changed $U_{tot}$ to $U_G$ in the gbond section to make the distinction between bond (Utot) and gbond (UG). you can change it as you see fit]}
\label{Fig:schematic_GBoNd}
\end{figure}
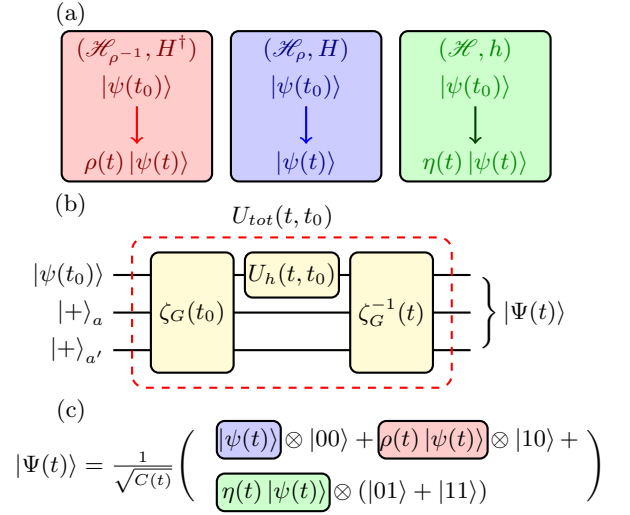 

\begin{figure*}[!t]
    \centering
    \includegraphics[width=1\textwidth]{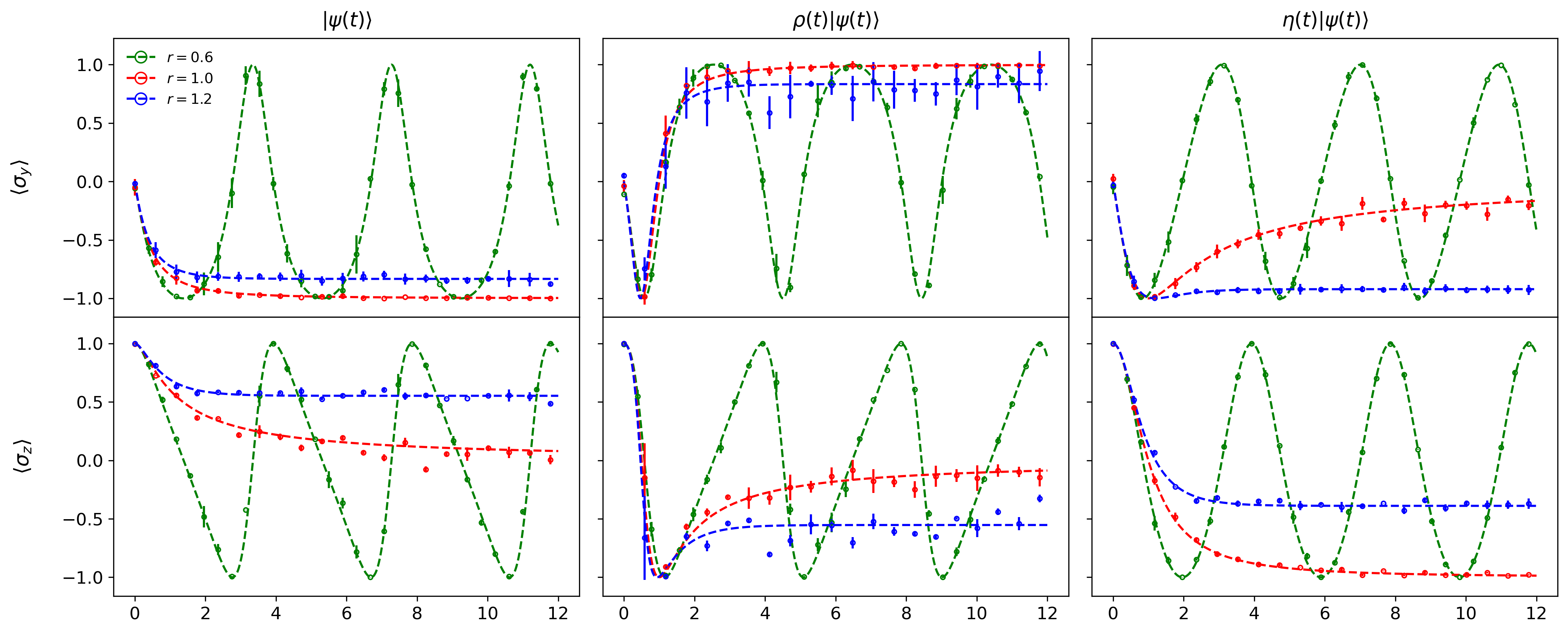}
    \caption{The spin expectation values  $\left<\sigma_y\right>$ and $\left<\sigma_z\right>$  calculated using the  time-evolved states $\ket{\psi(t)}/|\psi(t)|$, $\rho(t)\ket{\psi(t)}/|\rho(t)\psi(t)|$ (corresponding to the norm method with $H$ and $H^\dagger$, respectively), and  $\eta(t)\ket{\psi(t)}$ (representing the metric formalism).  The states are evolved using the GBoNd protocol with  the non-Hermitian Hamiltonian defined in Eq.\,\eqref{eqn:ham_exp} and the initial condition $\ket{\psi(t_0)}=\ket{0}$. The time is in units of [s]=1, and does not equate to physical time in the quantum circuit (see\,\cite{supplementmater}). The dots and the corresponding error bars are measurement results from the IBM digital quantum  device,  while the dashed lines are the analytical solutions. For our Hamiltonian,    $\left<\sigma_x\right>=0$ in all three states at all times.}
    \label{Fig:GBoNd_results}
\end{figure*}
% {\color{blue}  Need to define the expectation values that are plotted in the figures ...and highlight them accordingly}

%Our goal is to simulate the physics encapsulated by metric, in order to enable the experimental observation of intrinsic non-Hermitian dynamics.
To circumvent the aforementioned issues, we present a new dilation scheme which permits the direct implementation of both the norm method and metric formulation of non-Hermitian dynamics.
To this end, we note that 
\begin{equation}
%\begin{split}
        U_H(t,t_0)=\eta^{-1}(t)U_h(t,t_0)\eta(t_0), 
%    U_{H^\dagger}(t,t_0)&=\eta(t)U_h(t,t_0)\eta^{-1}(t_0),\\
 %   &=\rho(t)U_H(t,t_0)\rho^{-1}(t_0),
   % \end{split}
    \label{eq:evolmappingHh}
\end{equation}
where $U_h(t,t_0)$ is the time propagator corresponding to $h(t)$ given in Eq.\,\eqref{eq:H_h_mapp}. 
Since $\eta(t)$ is the only non-unitary operation in Eq.\,\eqref{eq:evolmappingHh}, we can specifically define a dilation on $\eta(t)$. In other words, labelling the general dilation of $\eta(t)$ as $\zeta(t)$, we effectively implement
\begin{equation}
    U_{tot}(t,t_0)=\zeta^{-1}(t)(U_h(t,t_0)\otimes\mathbb{I}_a)\zeta(t_0),
\label{eq:general_dil_scheme}
\end{equation}
where the second system is a two-level ancilla  belonging to the Hilbert space $\mathscr{A}\ni\mathbb{I}_a$. 
The Hamiltonian associated to Eq.\,\eqref{eq:general_dil_scheme} is given by $H_{tot}(t)=\zeta^{-1}(t)(h(t)\otimes\mathbb{I}_a)\zeta(t)+i\dot{\zeta}^{-1}(t)\zeta(t)$, with the Hilbert space  given by $(\h\otimes\mathscr{A},H_{tot}(t))$.
A general form  of the dilation $\zeta^{-1}(t)$ in Eq.\,\eqref{eq:general_dil_scheme} is given by
%In general, we can define a non-unique dilation of $\eta(t)$ with one ancilla qubit as 
% \textcolor{red}{(Maybe add the general state $\frac{1}{\sqrt{C(t)}}\left(\ket{\psi(t)}+\sqrt{C(t)\rho(t)-\mathbb{I}}\ket{\psi(t)}\right)$ inline?)}
{\fontsize{8.5pt}{10pt}
\begin{align}
\zeta^{-1}_{C(t)}(t)=\frac{1}{\sqrt{C(t)}}
\begin{bmatrix}
\eta^{-1}(t) & \sqrt{C(t)\mathbb{I} - \rho(t)^{-1}} \\
\sqrt{C(t)\mathbb{I} - \rho^{-1}(t)} & -\eta^{-1}(t)
\end{bmatrix},
\label{eq:general}
\end{align}
}
 where $C(t)$ is a system-dependent real scalar function which can be chosen to ensure that the dilation is always well defined. 
 We note that this dilation is not unique.
With the choice $C(t)=1$ in Eq.\,\eqref{eq:general}, we recover the Naimark dilation. 
% $$\zeta_N^{-1}(t)=
% \begin{bmatrix}
% \eta^{-1}(t) & \sqrt{\mathbb{I} - \rho(t)^{-1}} \\
% \sqrt{\mathbb{I} - \rho(t)^{-1}} & -\eta^{-1}(t)
% \end{bmatrix},$$ 
For our goals, an apt choice for two-level systems is $C(t)\propto tr[\rho(t)+\rho^{-1}(t)]$, which ensures the validity of our dilation at all times and significantly simplifies the implementation.

\textit{Generalized Biorthogonal Naimark dilation (GBoNd).}---Having defined a general dilation of $\eta(t)$ to implement non-unitary dynamics, we now turn to a scheme which allows us to access both interpretations $(\hrho, H(t))$ and $(\h, h(t))$ in the metric formalism.  Ideally, we want to obtain a final system-ancilla state of the form
\begin{equation}
\ket{\Psi(t)}\propto\ket{\psi(t)}_s\otimes\ket{0}_a+\eta(t)\ket{\psi(t)}_s\otimes\ket{1}_a,
\end{equation}
where $\ket{\psi(t)}$ and $\eta(t)$ are as defined previously. 
This implies that the time evolution of $\ket{\Psi(t)}$, which spans the Hilbert space$(\h\otimes\mathscr{A},H_{s,a}(t))$, is governed by a non-Hermitian total Hamiltonian $H_{s,a}(t)=H(t)\otimes\ket{0}\bra{0}_a+h(t)\otimes\ket{1}\bra{1}_a$, where $H(t)$ is the non-Hermitian Hamiltonian of interest and $h(t)$ is the Hermitian counterpart as introduced in Eq.\,(\ref{eq:H_h_mapp}). Consequently, we need an additional dilation to be able to realize the desired time evolution with unitary gates.

To this end, we introduce the metric operator $\rho_G(t)=\rho(t)\otimes\ket{0}\bra{0}_a+\mathbb{I}\otimes\ket{1}\bra{1}_a$, where $\rho_G(t)$ and $\rho(t)$ are the solutions of Eq.\,\eqref{eq:tdqh} corresponding to $H_{s,a}(t)$ and $H(t)$, respectively. The principal root of $\rho_G(t)$ is given by $\eta_G(t)=\eta(t)\otimes\ket{0}\bra{0}_a+\mathbb{I}\otimes\ket{1}\bra{1}_a$.
We now add a second ancilla $\mathscr{B}$ to obtain the dilation of $\eta_G(t)$, given by Eq.\,\eqref{eq:general}. In the rest of the paper, we focus on two-level systems and set $C(t)=tr[\rho_G(t)+\rho_G^{-1}(t)]D^{-1}$, where $D=tr[\mathbb{I}_{s,a}]$. With this choice, the expression for the dilation simplifies to
\begin{equation}
    \zeta_G^{-1}(t)=\frac{1}{\sqrt{C(t)}}
\begin{bmatrix}
\eta_G^{-1}(t) & \eta_G(t) \\
\eta_G(t) & -\eta_G^{-1}(t)
\end{bmatrix},
\label{eqn:zetag}
\end{equation}
where $D$ is the dimension of the non-Hermitian Hilbert space $\mathscr{H}$ and $\zeta_G(t)$ is now defined in the Hilbert space $\h\otimes\mathscr{A}\otimes\mathscr{B}$. The total time evolution  operator  is now given by $U_{tot}(t,t_0)=\zeta_G^{-1}(t)(U_h(t,t_0)\otimes\mathbb{I}_{a,b})\zeta_G(t_0)$. The implementation of this dilation is schematically shown in Fig. \ref{Fig:schematic_GBoNd}. 
Considering an initial state $\ket{\Psi(t_0)}=\ket{\psi(t_0)}_s\otimes\ket{+}_a\otimes\ket{+}_{b}$, where $\ket{+}=\frac{1}{\sqrt{2}}(\ket{0}+\ket{1})$, the final state $\ket{\Psi(t)}=U_{tot}(t,t_0)\ket{\Psi(t_0)}$ implementing our GBoNd protocol is given by
\begin{equation}
\begin{split}
\ket{\Psi(t)}=\frac{1}{\sqrt{C(t)}}\bigg(&\ket{\psi(t)}_s\ket{00}
+\rho(t)\ket{\psi(t)}_s\ket{10}\\
&+\eta(t)\ket{\psi(t)}_s(\ket{11}+\ket{01})\bigg),
\end{split}\label{eq:U_g_state}
\end{equation}
In the above equation, for the sake of readability, we have dropped the subscripts $\ket{ij}_{a,b}:=\ket{ij}$ and the tensor product.

Our dilation scheme encompasses the full array of non-Hermitian physics: the norm method  as well as the self-consistent metric formulation.
Specifically, the different projections of the ancillas provide a direct probabilistic access to time-evolved states corresponding to the Hamiltonians $H(t)$, $H^\dagger(t)$ and $h(t)$, namely $\ket{\psi(t)}$, $\rho(t)\ket{\psi(t)}$ and $\eta(t) \ket{\psi(t)}$.
% and observables in these states. 
Note that the first two correspond to the right and left states in biorthogonal quantum mechanics\,\cite{brody2013biorthogonal}. 
%The final wavefunction allows for inherently probabilistic access to the different states which results from the inherent need of a projection operator to observe non-unitarity.
% {\color{blue} $H_G$ not defined ..}
A particular advantage of our dilation protocol is that via $\ket{\psi(t)}$ and $\rho(t)\ket{\psi(t)}$, we can directly measure the dynamical metric operator $\rho(t)$ using state tomography on the single states in post-selection (see\,\cite{supplementmater}). 
% This in turn allows for a self-contained discrete evolution in which we redefine the total Hamiltonian at each time slice after the measurement of the metric operator $\rho(t)$, in a feedback-loop-like fashion. 
% \textcolor{olive}{This unveils a deeper connection between the metric dynamics and its role in ensuring the dynamical structure of the Hilbert space \((\hrho, H(t))\). 
% For further details on the self contained evolution scheme, we refer to the Supplementary Text\,\cite{supplementmater}}.
%{\color{blue} We need to mention tomography etc as a way to observe $\eta$ and observables which show the true inner product... we need a few comments on the projective measurements, whether this is deterministic etc etc }

We now implement the two-ancilla GBoNd protocol in a quantum circuit and provide the first experimental realization of the metric formalism and its dynamical inner product.
To this end, we consider a single two-level system described by the non-Hermitian Hamiltonian 
\begin{equation}
    H=\sigma_x+ir\sigma_z.
    \label{eqn:ham_exp}
\end{equation}
% \textcolor{blue}{(We set Gamma to 1 throughout right? Maybe just remove the Gamma parameter in the equation, text and figures? Otherwise modify the text below.)}
This Hamiltonian has a spectrum $\pm\sqrt{1 - r^2}$, which hosts exceptional points at $r=\pm1$. For $|r|>1$, the $\mathcal{PT}$-symmetry is spontaneously broken and the spectrum consists of complex conjugate pairs.
For further details on the Hermitian counterpart $h(t)$ we refer to\,\cite{supplementmater}.

The total time propagator $U_{tot}(t,t_0)$, as shown in Fig.\,\ref{Fig:schematic_GBoNd}, can be realized with the help of an approximate quantum compiler\,\cite{madden2022best,supplementmater} in a circuit layout as a sequence of universal gates and rotations. 
Our simulation is performed on the quantum processing unit \textbf{ibm\_Kyiv} provided by the IBM. We choose a circuit depth of $3$ to approximate the evolution operator $U_{tot}(t,t_0)$ and use gate-twirling options to minimize errors. 
A minimum of $N=10^3$ shots was considered to extract the states corresponding to the different dynamics $H$, $H^\dagger $ and $h(t)$. However, due to the probabilistic nature of our protocol, the true number of shots used is $N_{eff}=N/p_{0}$, where $p_0$ is the lowest projection probability corresponding to the desired ancilla configuration. 
For further details, we refer to the Supplemental Material\,\cite{supplementmater}.

We focus on the three spin expectation values defined by
\begin{equation}
    \left<\sigma_i\right>=\frac{\bra{\chi(t)}\sigma_i\ket{\chi(t)}}{\left<\chi(t)|\chi(t)\right>},
\end{equation}
where $\ket{\chi(t)}\in\{\ket{\psi(t)},\rho(t)\ket{\psi(t)}, \eta\ket{\psi(t)}\}$ and $i\in\{x, y, z\}.$
Our hardware simulation results for these spin expectation values are shown in Fig.\,\ref{Fig:GBoNd_results}.
Note that the three states manifest different behaviors as expected. The expectation values
evaluated in $\ket{\psi(t)}$ represents the norm method valid for dissipative systems, while that with $\eta(t)\ket{\psi(t)}$ represents the self-consistent metric formalism. 
We see that in all three states, the $\mathcal{PT}$-symmetric regime is characterized by oscillatory behavior while asymptotic behaviors are observed in the $\mathcal{PT}$-broken regimes. 
Clearly, the GBoNd protocol easily accesses the $\mathcal{PT}$-broken regime $|r|>1$, making it more versatile than the Naimark dilation\,\cite{beneduci2020notes, naimark_science,dogra2021quantum}. 
In both regimes, we see that the observables calculated in the metric formalism exhibit qualitatively different behaviours when compared to those using the norm method. In particular, at the 
exceptional points $r=\pm 1$, $\left<\sigma_y\right> \to 0$ and $\left<\sigma_z\right> \to -1$ in the metric formalism, while the inverse happens in the norm method.
 The fundamentally different physical behaviors predicted by the two approaches highlights the possibility of novel phenomena in the self-consistent non-Hermitian realm.

\textit{Biorthogonal Naimark dilation (BoNd).}---Having demonstrated a general dilation protocol using two ancillas, we turn to a simplified, one-ancilla scheme based on a dilation of $\eta(t)$. To this end, we replace $\eta_G(t)$ by $\eta(t)$ in Eq.\,\eqref{eqn:zetag} and set $C(t)=tr[\rho(t)+\rho^{-1}(t)]D^{-1}$, where $D=tr[\mathbb{I}_s]$. This allows us to define the total time propagator analogously to GBoNd, $U_{tot}(t,t_0)=\zeta^{-1}(t)(U_h(t,t_0)\otimes\mathbb{I})\zeta(t_0)$. 
Taking the initial state $\ket{\Psi(t_0)}=\ket{\psi(t_0)}_s\otimes\frac{1}{\sqrt{2}}(\ket{0}+\ket{1})$, we have the time-evolved state 
%$\ket{\Psi(t)}=U_{tot}(t,t_0)\ket{\Psi(t_0)}$ is
\begin{equation}
    \ket{\Psi(t)}=\frac{1}{\sqrt{C(t)}}(\ket{\psi(t)}_s\ket{0}+\rho(t)\ket{\psi(t)}_s\ket{1}).
    \label{eq:BoNdstate}
\end{equation}
This dilation encompasses the states  $\ket{\psi(t)}$ and $\rho(t)\ket{\psi(t)}$, which evolve with the Hamiltonians $H$ and $H^\dagger$, respectively. They correspond to the right and left states in biorthogonal quantum mechanics\,\cite{brody2013biorthogonal}.
Similarly to GBoNd, the BoNd protocol enables the access to the $\mathcal{PT}$-broken regime for arbitrary times. 
This is contrasted with the Naimark dilation\,\cite{paulsen2002completely, naimark_science,dogra2021quantum}, which is ill-defined for arbitrary times, since the total dilated Hamiltonian contains divergent terms like $\tilde{\eta}^{-1}(t)$. 
Ref.\,\cite{dogra2021quantum} worked around the problem by redefining the  prefactor $C(t)$ at every time step in the $\mathcal{PT}$-broken regime. Therefore, the protocol samples $\ket{\psi(t)}$ from multiple total dilations, which means the total protocol is not a single Hermitian evolution. %\textcolor{blue}{(check: single Hermitian or non-Hermitian evolution?)} \textcolor{olive}{It is multiple Hermitian dilation protocols} 
% \textcolor{blue}{Maybe if we can find the paper on impossible to simulate $PT$-symmetry broken regimes we can add it here as a comment too.}

\begin{figure}[!t]
    \centering
    %Label
    % \begin{minipage}[t]{0.235\textwidth}
    %     \centering
    %     \begin{tikzpicture}
    %         \node at (2.5,2.9) {\includegraphics[width=0.6\linewidth]{images/paper_thesis_figures_final/legend_fig_3.png}};
    %     \end{tikzpicture}
    % \end{minipage}
    % \\
    
    % First row
    \begin{minipage}[t]{0.238\textwidth}
        \centering
        \begin{tikzpicture}
            \node at (0,0) {\includegraphics[width=\linewidth]{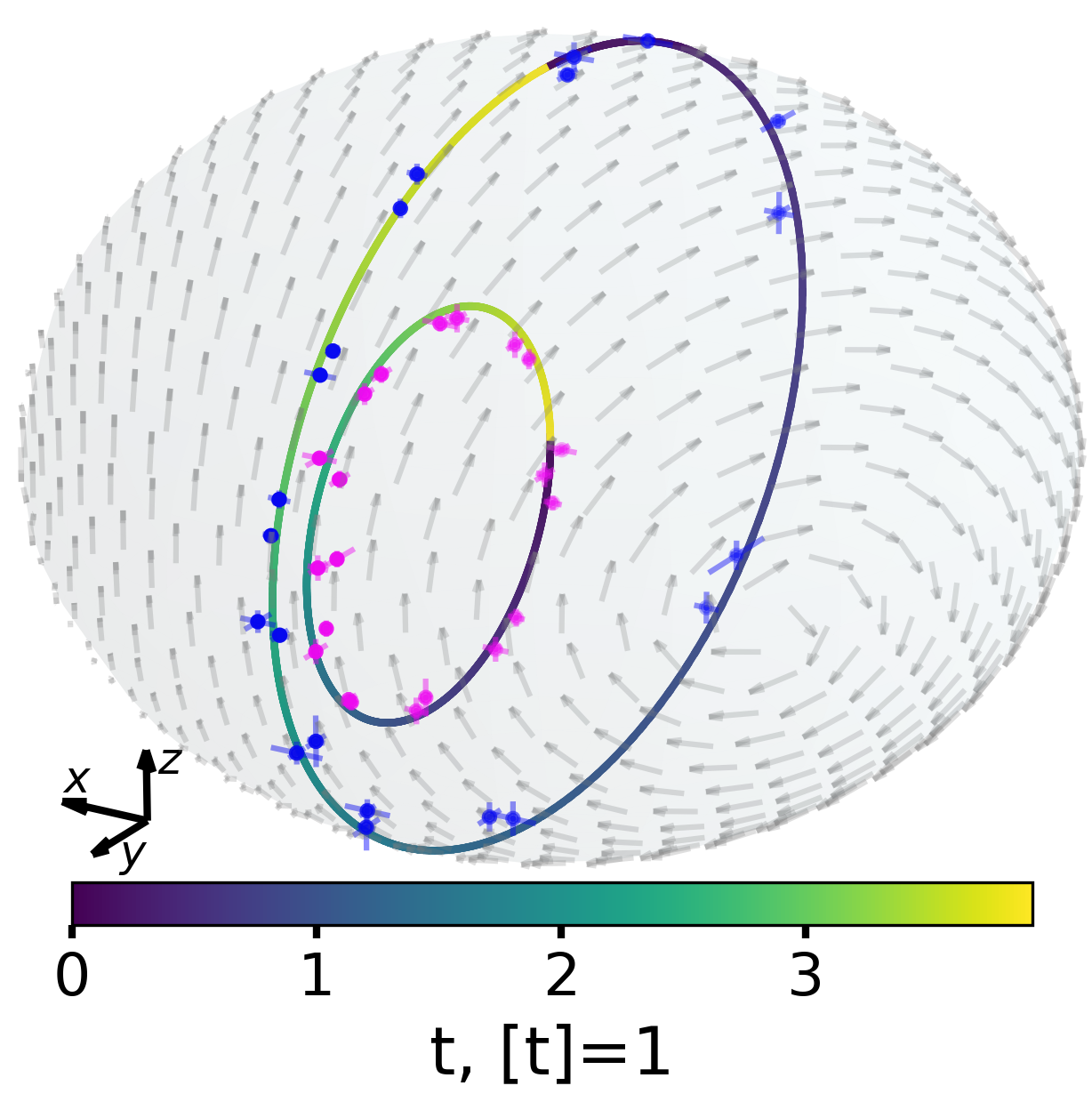}};
            \node[anchor=north west] at (-2.3,2.2) {\textbf{(a)}};
        \end{tikzpicture}
    \end{minipage}
    % \hfill
    \begin{minipage}[t]{0.238\textwidth}
        \centering
        \begin{tikzpicture}
            \node at (0,0) {\includegraphics[width=\linewidth]{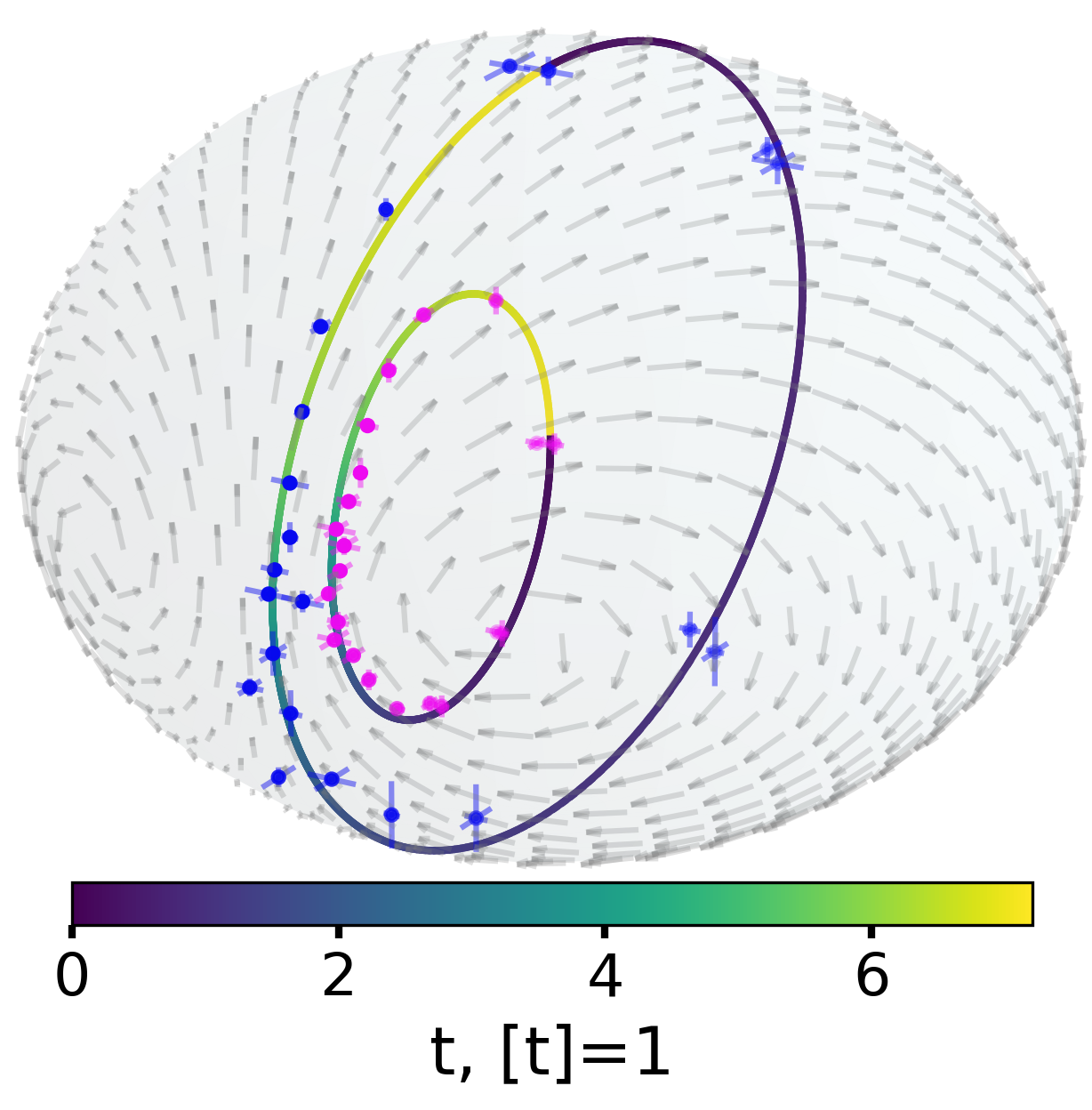}};
            \node[anchor=north west] at (-2.3,2.2) {\textbf{(b)}};
        \end{tikzpicture}
    \end{minipage}
    \\
    \vspace{0.5em}

    % Second row
    \begin{minipage}[t]{0.238\textwidth}
        \centering
        \begin{tikzpicture}
            \node at (0,0) {\includegraphics[width=\linewidth]{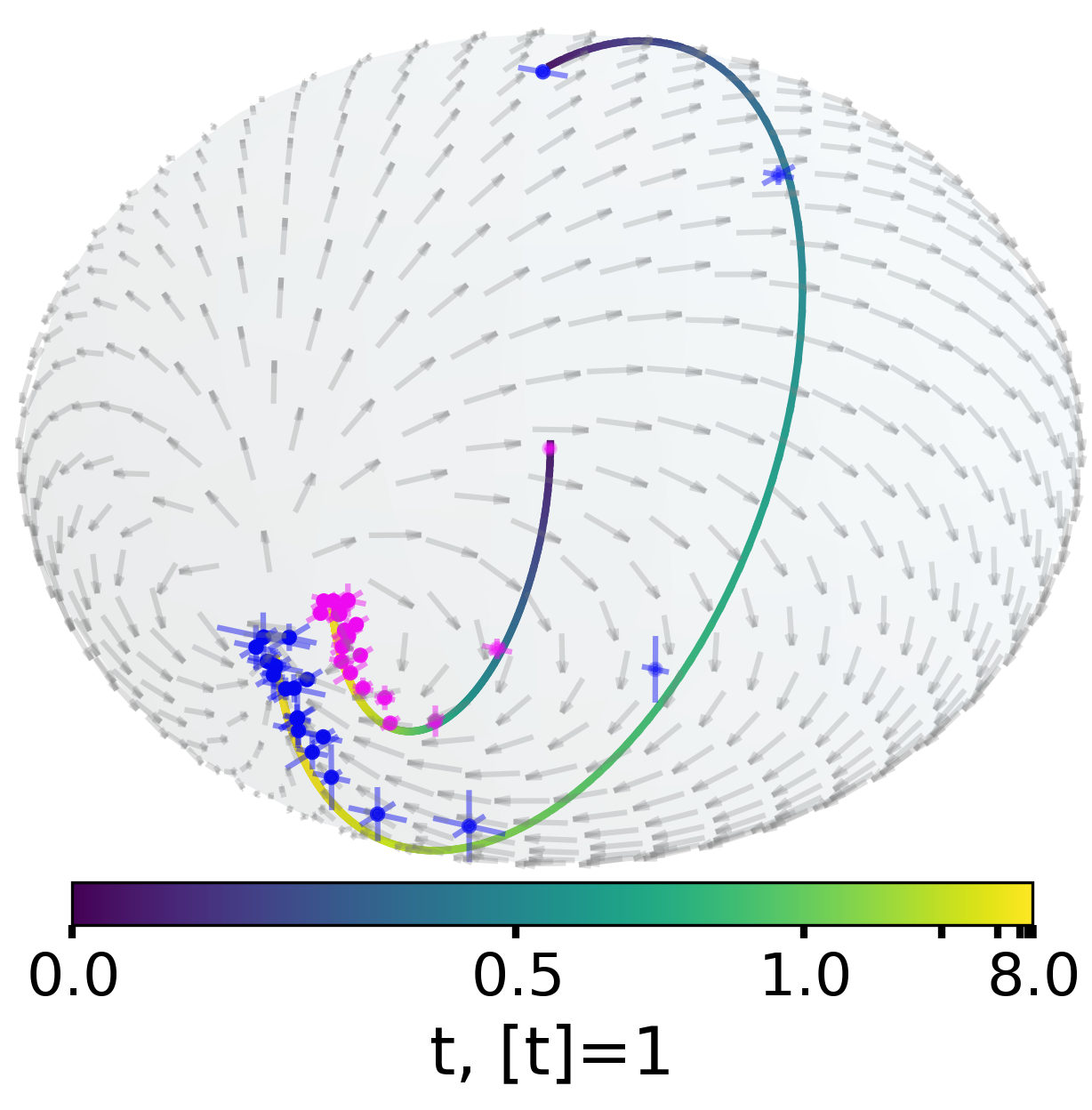}};
            \node[anchor=north west] at (-2.3,2.2) {\textbf{(c)}};
            \node at (0.7,-2.4) {\includegraphics[width=0.6\linewidth]{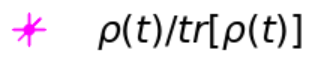}};
        \end{tikzpicture}
    \end{minipage}
    % \hfill
    \begin{minipage}[t]{0.238\textwidth}
        \centering
        \begin{tikzpicture}
            \node at (0,0) {\includegraphics[width=\linewidth]{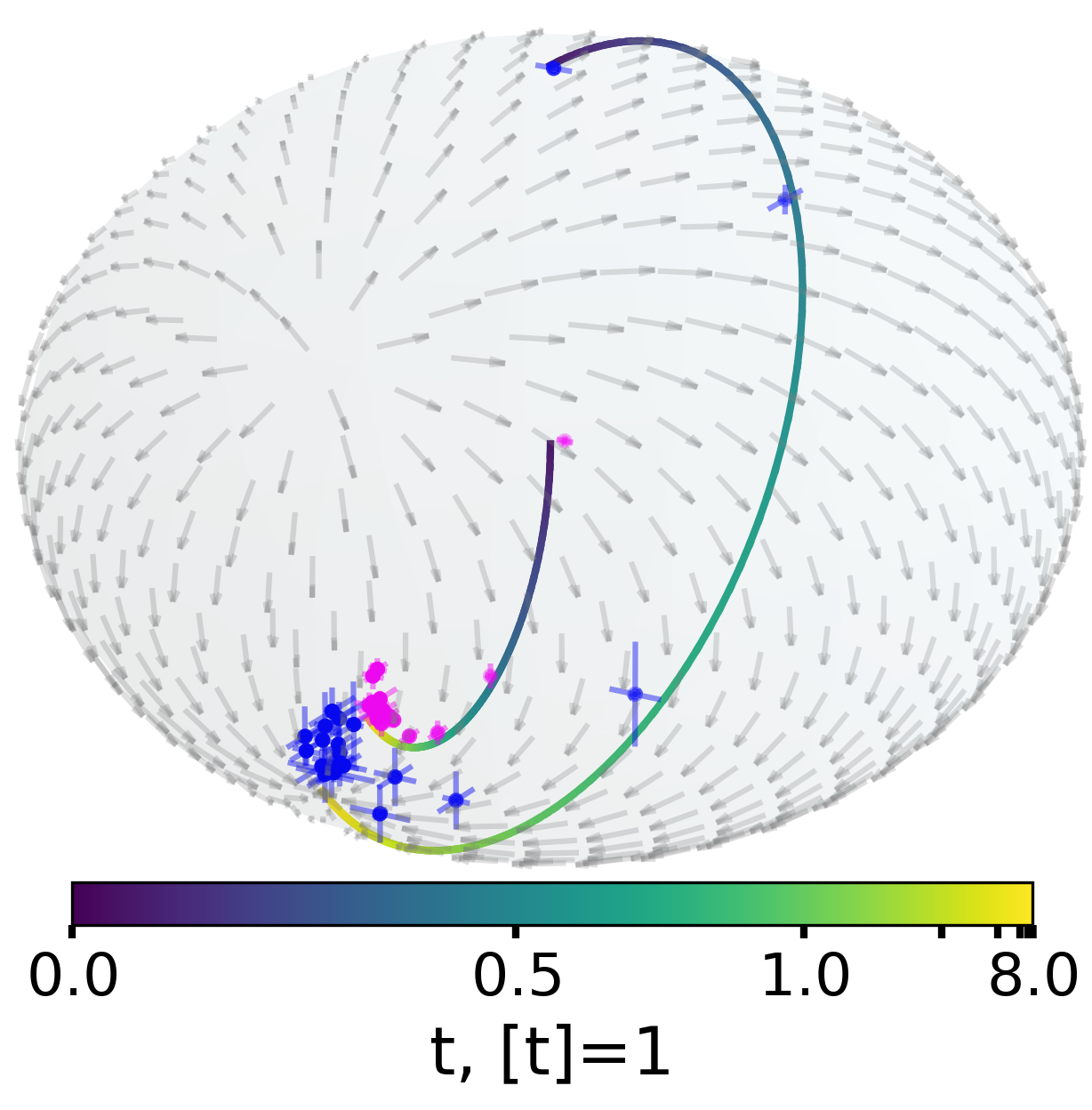}};
            \node[anchor=north west] at (-2.3,2.2) {\textbf{(d)}};
            \node at (-0.7,-2.4) {\includegraphics[width=0.6\linewidth]{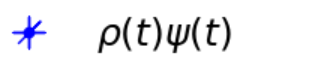}};
        \end{tikzpicture}
    \end{minipage}

    \caption{The evolutions of the normalized state $\rho(t)\ket{\psi(t)}/|\rho(t)\psi(t)|$ and  metric $\rho(t)/tr[\rho(t)]$ under the non-Hermitian Hamiltonian $H^\dagger=\sigma_x-ir\sigma_z$, as defined in Eq.\,\eqref{eqn:ham_exp}, with (a) $r=0.6$, (b) $r=0.9$, (c) $r=1$, (d) $r=1.2$. The vector field points towards $(\partial_t\ket{\chi(t)}\bra{\chi(t)})dt$, where $\ket{\chi(t)}$ is a pure state on the Bloch sphere. The points are measurement results obtained from a physical circuit using the BoNd protocol, while the lines are the analytical solutions. The metric components were multiplied by a factor $f=0.8$ for $r\geq0.9$ to distinguish the two lines.
    % \textcolor{red}{The points are measured and the lines are analytical results?}
    }
    \label{Fig:BoNd_results}
\end{figure}
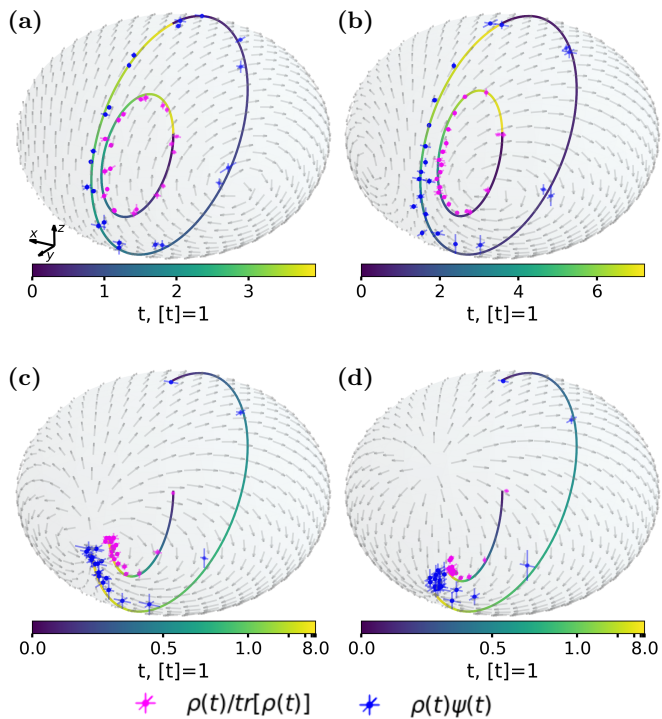
To directly obtain $\rho(t)$, we implement the BoNd protocol for the Hamiltonian $H$ in Eq.\,\eqref{eqn:ham_exp} with the same circuit setup used for GBoNd. 
Using state tomography on the wavefunction in Eq.(\ref{eq:BoNdstate}), we obtain the states $\ket{\psi(t)}$, $\rho(t)\ket{\psi(t)}$ and the operator $\rho(t)/tr[\rho(t)]$. The latter two are shown in Fig.\,\ref{Fig:BoNd_results}. As with GBoNd, there is good agreement with the analytical results in both $\mathcal{PT}$-symmetric and $\mathcal{PT}$-broken regimes. However, note that the BoNd does not permit the direct measurement of the observables within the metric formalism, contrary to GBoNd.

Figure \ref{Fig:BoNd_results} highlights an important aspect of the metric $\rho(t)$: it is time-periodic in 
the $\mathcal{PT}$-symmetric regime.
Defining $\rho_C=\oint\rho(t)dt=e^{\beta\sigma_y},\,\tanh(\beta)=r$, where $\oint$ denotes the time average over a period, we see that $\oint(H^\dagger\rho(t)-\rho(t)H)dt=\oint i\frac{d}{dt}\rho(t)dt=0$. 
This implies that $\rho_C$, a stationary solution of Eq.\,\eqref{eq:tdqh}, defines the pseudo-Hermiticity similarity transformation $H^\dagger\rho_C=\rho_CH $ expected in the $\mathcal{PT}$-symmetric region of a constant Hamiltonian\,\cite{mostafazadeh2010pseudo} %\textcolor{blue}{(add ref; possibly Mostafazadeh. ideally one that uses the term ``pseudo-Hermiticity" to refer to the similarity transformation)}\textcolor{olive}{only one? I have like 4 earlier above}.
This reiterates that the dynamical metric encodes the stationary similarity transformation, making it the self-consistent approach for generic non-Hermitian Hamiltonians. Note that for general
 Hamiltonians $H$ in the $\mathcal{PT}$-symmetric regime, the time propagator is typically quasi-periodic. This also requires the usage of the more general dilation scheme given in Eq.\,\eqref{eq:general}, where the choice of $C(t)$ will be system specific.
Nonetheless, the quasiperiodicity can be used to define $\rho'_C=\int_{t_0}^{t_1}dt\rho(t)$ where $t_1 >> t_0$, such that $|H^\dagger\rho'_C-\rho'_CH|=|\rho(t_0)-\rho(t_1)|<\epsilon$ for some appropriately small bound $\epsilon>0$.

% \textcolor{blue}{This is already written in an earlier passage, but it can be reiterated in the conclusion}\textcolor{purple}{Another advantage of the BoNd and GBoNd dilation scheme is the self-contained evolution that this dilation offers.  They unveil deeper connections between the metric in the Hilbert space and .....
% In the supplementary material\,\cite{supplementmater}, we show that the time evolution can be performed by extracting the metric from the state at each timestep and updating the total Hamiltonian based on the measured metric. 
% This is analogous to general relativity, where the energy density matrix and spacetime dictate their mutual evolution.}\\

\textit{Conclusion.}---We show, via  the use of   generalized dilation protocols, how  the different formulations of non-Hermitian quantum mechanics  can be simultaneously accessed in  larger closed systems undergoing purely unitary dynamics.  Implementing our protocols on a digital simulator,  we presented the first experimental demonstration of non-Hermitian quantum mechanics characterized by non-stationary Hilbert spaces.
Specifically, our results unfurl the  self-consistent dynamics of the metric operator  which endows the Hilbert space with a dynamical inner product. 
%An advantage of our dilation schemes is the self contained evolution that it offers, which relies on the fact that the metric $\rho(t)$ can be measured at each time point through state tomography of the wavefunction after the projection of the ancillas (Eq. (\ref{eq:U_g_state}) and Eq. (\ref{eq:BoNdstate})). 
As we briefly discuss in the Supplemental Material\,\cite{supplementmater},  the interplay between the time-evolving metric and the  updated   total Hamiltonian dictating the dynamics of the dilation  has  structural similarities to  what is seen in  general relativity, where the energy-momentum tensor and spacetime dictate their mutual evolution.  The theoretical dilation framework developed in this work is promising for further studies of non-Hermiticity, including the non-Hermitian skin effect\,\cite{molignini2023anomalous}, non-Hermitian field theory\,\cite{mostafazadeh2006physical}, as well as the geometry of dynamical Hilbert spaces through the Fubini-study metric analogous to Ref.\,\cite{mostafazadeh2007quantum}. 
Our work provides a new impetus to the field of non-Hermitian quantum physics, paving the way for the discovery of unconventional phenomena.

\bibliography{apssamp}% Produces the bibliography via BibTeX.

\end{document}